\documentclass[11pt,a4paper]{article}
\usepackage{amsmath,amssymb}
\usepackage{epsfig,graphicx}

\def\be{\begin{equation}}
\def\ee{\end{equation}}
\def\bea{\begin{eqnarray}}
\def\eea{\end{eqnarray}}

\def\beq{\begin{equation}}
\def\eeq{\end{equation}}

\begin{document}

\title{\bf Shrinking fermionic modes,\\ on the lattice and in the continuum}
\author{V.~I.~ Zakharov$^{a,b}$\footnote{{\bf e-mail}: xxz@mppmu.mpg.de}
\\
$^{a}$ \small{\em Istituto Nazionalr  di Fisica Nucleare - Sezione di Pisa
} \\
\small{\em Largo Pontecorvo, 3, 56127 Pisa, Italy }\\
$^{b}$ \small{\em Institute of Theoretical and  Experimental Physics}\\
\small{\em B.~Cheremushkinskaya~25, Moscow, 117259, Russia }}
\date{}
\maketitle

\begin{abstract}
Recent lattice data indicates that volume occupied by
topological fermionic modes shrinks to zero in the continuum limit
of vanishing lattice spacing. The data apparently cannot be accommodated 
within, say, conventional instanton model. We review field-theoretic arguments 
which demonstrate that the topological fermionic modes are to shrink
to a vanishing submanifold of the whole four dimensional space
provided that  measurements are performed with high resolution.
Moreover, the data fit well the emerging  overall picture of lower-dimensional
defects in the Euclidean vacuum of Yang-Mills theories.
We also mention results on localization of scalar particles.
\end{abstract}

\section{Introduction}

In this talk we will discuss topological defects in the Euclidean Yang-Mills vacuum.
By topological defects we understand regions with large 
absolute value of 
the density of the topological charge $Q_{top}(x)$,
\begin{equation*}
 Q_{top}(x)~=~(16\pi^{2})^{-1}G^{a}_{\mu\nu}(x)\tilde{G}^{a}_{\mu\nu}(x).
\end{equation*}
 Usually one thinks about such regions in terms of instantons.
For instantons, \footnote{The instanton model has been elaborated
in great detail, for review see \cite{shuryak,teper1}.}
\begin{equation*}
\Big(\int d^{4}xQ_{top}(x)\Big)_{instanton}~=~1~~,
\end{equation*}
which is large compared to the perturbative noise which
contains an extra power of $\alpha_{s}^{2}$.
The instanton picture has been  challenged
since long because of inconsistencies in the large $N_{c}$ limit
\cite{witten3,horvath}.
An alternative description 
could be provided by
domain walls \cite{witten3}. The theory of domain walls is not developed in detail, however. 

Note that domain walls are, by definition, three-dimensional (3d) defects 
in the vacuum.
Thus, one could argue that in the domain-walls picture topological defects would occupy a vanishing fraction
of the whole 4d space. 
As far as we know, however, this point has never been emphasized.
Moreover,  the first example
of low-dimensional vacuum defects 
was provided by lattice strings, or vortices
\footnote{For review of the role of the vortices in the confinement see \cite{greensite} while identification
of the vortices with strings is based on results of papers
\cite{ultraviolet} and introduced  in \cite{vz5}.}
which are responsible for the confinement. 

The possibility that topological defects could occupy a
3d submanifold of the 4d space was suggested first in Ref \cite{vz5}
in the context of the 3d defects discovered in \cite{3d}
and closely related to the vortices \footnote{No suggestion was made, however,
how to verify this prediction on the lattice. Moreover, even now, that 
there is emerging data that topological defects are indeed shrinking to
a vanishing, three-dimensional submanifold of the whole space,
see below, there is no proof that this submanifold is the same
as suggested in Ref. \cite{vz5}. Nevertheless, it is amusing
that such a simple classification scheme of the defects as presented
in \cite{vz5} could determine the dimension of the chiral defects
correctly. }. 
Independently, there began to appear data on unusual behavior
of fermionic zero modes as function of the lattice spacing
\cite{random}.
The data does indicate that topological defects shrink to a vanishing subspace of the whole
space if measurements are performed with high resolution,
that is on scale of the lattice spacing.

\section{Low-lying fermionic modes}
\subsection{Generalities}
To uncover topology of the gluonic fields one concentrates  on 
low-lying modes
of the Dirac operator.  The modes are defined as solutions of the eigenvalue
problem
\begin{equation}\label{lambda}
D_{{\mu}}\gamma_{\mu} \psi_\lambda \, =\, \lambda \psi_\lambda\;,
\end{equation}
where the covariant derivative  is constructed on the vacuum
gluonic fields  $\{A_{\mu}^a(x)\}$. 

For exact zero modes,
the difference between the number of
modes with positive and negative chirality equals to the total
topological charge of the lattice volume:
\begin{equation}\label{zeromodes}
n_{+}-n_{-}~=~Q_{top}\;.
\end{equation}
Assuming that topological charge fluctuates 
by order unit
independently on pieces of  4d volumes measured in physical units
one derives:
\beq \label{topologicalcharge}
\langle Q_{top}^{2}\rangle~\sim ~\Lambda_{QCD}^{-4}V_{tot}~
\approx(180MeV)^{4} V_{tot}~~~,
\end{equation}
where the numerical coefficient here is related to the $\eta^{'}$-mass
(the Witten-Veneziano relation).

One also considers so called near-zero modes which occupy, roughly speaking
the interval
\begin{equation}\label{band}
0~~<~~\lambda~~<~~{\pi\over L_{latt}}~~,
\end{equation}
where $L_{latt}$ is the linear size of the lattice.
Near-zero modes determine the value of the quark condensate via
the Banks-Casher relation:
\begin{equation}\label{bankscasher}
\langle \bar{q}q \rangle ~=~ -\pi \rho(\lambda \to 0)\;,
\end{equation}
where  $\lambda \to 0$ with the total volume tending to infinity.

\subsection{Lattice data}

While the close  relation of the low-lying fermionic modes to the topology
of the underlying gluon fields is well known since long,
it is only  recently that   these modes have been measured on the
original field configurations, without cooling.
This recent progress  is due to  the advent of the overlap operator \cite{neuberger}.

Measurements  on original fields \cite{random}
confirmed the general relations (\ref{topologicalcharge})
and (\ref{bankscasher}). However, they also 
 brought an unexpected result
that the volume occupied
by low-lying modes apparently tends to zero in the continuum limit of
vanishing lattice spacing, $a\to 0$. Namely,
\begin{equation}\label{shrinking}
\lim_{a\to 0}{V_{mode}}~\sim~(a\cdot \Lambda_{QCD})^{r}~\to ~0\;,
\end{equation}
where $r$ is a positive number of order unit and the volume occupied by a mode,
$V_{mode}$ is defined in terms of the Inverse Participation
Ratio (IPR) \footnote{Independent evidence in favor of shrinking
of the regions occupied  by topologically non-trivial gluon fields was
obtained in \cite{gubarev}.}.
As for exact numerical values of the critical exponent $r$,
one should address the original papers \cite{random} for details. 
Roughly speaking, measurements 
 mostly favor
$r\approx 1$, except for the second paper in Ref. \cite{random}
where $r$ is rather larger.
\subsection{Chirality and  the vortices}

A crucial  question is then, whether the underlying vacuum structure
is the same for the confining fields and fields with non-trivial
topology. 
An attempt to answer this question was undertaken in 
Ref. \cite{correlator}
through a direct  study of correlation between intensities of 
fermionic modes and of vortices.

In more detail,
center vortex is a set of plaquettes $\{D_i\}$ on the dual lattice,
for review see \cite{greensite}. Denote
the set of plaquettes dual to $\{D_i\}$ by $\{P_i\}$. Then the 
correlator in point is defined as:
\begin{eqnarray}\label{correlation}
C_\lambda(P) =
\frac{\sum_{P_i} \sum_{x \in P_i} (\rho_\lambda(x) -\langle \rho_\lambda(x) \rangle ) }
{\sum_{P_i} \sum_{x \in P_i} \langle \rho_\lambda(x) \rangle} \, ,
\label{eq:z2_plaq_corr_orig}
\end{eqnarray}
where $\rho_{\lambda}(x)$ is the density of the fermionic mode
with eigenvalue $\lambda$.  
Since $ \sum_x \rho_\lambda(x) = 1 $ and
$ \langle V_{tot} \rho_\lambda(x) \rangle = 1 $, definition (\ref{correlation}) can be rewritten as
\begin{eqnarray}
C_\lambda(P) = \frac{\sum_{P_i} \sum_{x \in P_i} (V_{tot}\rho_\lambda (x) - 1)}
{\sum_{P_i} \sum_{x \in P_i} 1}\,.
\label{eq:z2_plaq_corr}
\end{eqnarray}
Results of the measurements can be found in the original paper \cite{correlator}.
Here we   briefly summarize the findings.

First of all, there does exist positive correlation
between intensities of fermionic modes and
density of vortices. 
Second,  the value of the correlator 
depends on the eigenvalue
and the correlation is strong only for the topological fermionic modes.
Finally and most remarkably, the correlation grows with diminishing larttice
spacing. A simple  analysis reveals that, indeed, if the 2d defects 
are entirely responsible for chiral symmetry breaking or  
constitute a boundary of 3d defects carrying large topological charge,
the correlator (\ref{correlation}) grows as an inverse 
power of the lattice spacing. 
The data does show that the correlator grows for smaller $a$ but
does not allow yet to uniquely fix  the dimensionality of 
the chiral defects.

\section{Pieces of theory}
\subsection{``Subtraction volume''}
The result (\ref{shrinking}) is in striking contradiction with the instanton
model and at first sight seems very difficult to appreciate.
A more careful analysis demonstrates, however, that the shrinking of 
topological fermionic modes could have been predicted from field theory
\footnote{The argumentation was worked out by A.I. Vainshtein 
and the author \cite{arkady} and outlined in the talk \cite{adriano}.\label{alpha}}.

As is explained above the topological fermionic modes just reveal 
the topological structure of the underlying gluon fields and we can, therefore,
concentrate on distribution of $G\tilde{G}(x)$.
Consider correlator of the topological density  at two points.
From  general principles alone, one can  show that for any finite $x$
\footnote{For references and discussion see \cite{seiler}.}:
\begin{eqnarray}\label{general}
\langle~G\tilde{G}(x),G\tilde{G}(0)~\rangle_{Minkowski}~>~0\\
\langle~G\tilde{G}(x),G\tilde{G}(0)~\rangle_{Euclidean}~~<~0 ~.\nonumber
\end{eqnarray}
On the other hand, for a pure instanton, or within a  zero mode:
\beq\label{nonunitary}
\langle~G\tilde{G}(x),G\tilde{G}(0)~\rangle_{instanton}~~>~0~.
\end{equation}
Since the instanton contribution (\ref{nonunitary})
taken alone violates unitarity, compare to (\ref{general}),
it cannot dominate  and the unitarity is restored 
at any finite $x$ by perturbative contributions.
Somewhat schematically, the correlator can be represented 
at short distances as:
\beq\label{perturbative}
\langle~G\tilde{G}(x),G\tilde{G}(0)~\rangle_{Euclidean}~\sim~
-{c_{1}\alpha_{s}^{2}\over x^{8}}~+~c_{2}\Lambda_{QCD}^{4}\delta (x)~,
\end{equation}
where $c_{1,2}$ are positive constants and the $1/x^{8}$ term is perturbative.

The central point is that by measuring topological modes we filter 
the perturbative noise away and are left with the local term. 
In the language of dispersion relations, this 
is a subtraction term, which has no imaginary part \cite{seiler}.

It is only natural then that contributions which are described
by subtraction constants in dispersion relations appear
as vanishing submanifolds  once attempt is made to measure their
spatial extension, or volume. Moreover, to see that the volume is small
we need measurements with high resolution. Hence, dependence on
the lattice spacing exhibited by the data (\ref{shrinking}). 
In this sense, the lattice spacing $a$ is to be understood  as resolution
of the measurements rather than  an ultraviolet cut off.

\subsection{Dimensionality of the chiral defects}

Although this type of argument makes observation (\ref{shrinking})
absolutely natural and predictable, it does not immediately fix the exponent $r$.
Further considerations seem to favor $r=1$, or 3d defects. 
Here we briefly mention the arguments in favor of 3d topological defects
(see also \cite{vz5}). 

First, we already mentioned that domain walls appear naturally
in dual formulations of Yang-Mills theories with large $N_{c}$.
It is not a proof of $r=1$ yet since quantum corrections at $N_{c}=2,3$ 
could induce non-trivial fractal dimensions
so that 3d manifold would percolate through the 4d space and occupy
the whole or a finite part of it
\footnote{Such defects were considered in Ref. \cite{thacker}. 
Since these topological defects do not depend on  the $\Lambda_{QCD}$ scale, 
they are the same typical for Yang-Mills as, say, for photodynamics.
In these notes, we do not consider non-trivial fractal dimensions
and by 3d defects always understand 3d volumes in physical units
which are a vanishing part of the total volume in the continuum limit.}.
From experience
with the lattice strings, see, e.g. \cite{vz5}, we would still expect
zero anomalous fractal dimension and topological defects occupying 3d volume in physical units.

 More specifically, one can invoke analogy with quantum mechanics \cite{arkady}. 
 In case of quantum mechanics, the phenomenon is 
 that if one tries to measure the time spent by a particle under the barrier,
 this time turns to be zero. A mnemonic rule is that the particle under the barrier
 lives
 in  imaginary time  and when projected to real
 time the barrier transition has zero duration (for details 
 and references see \cite{adriano}). 
 Violation of the unitarity in (\ref{nonunitary}) could also be formally
 removed by changing one coordinate from real to imaginary values. 
 Thus, one expects that only one coordinate collapses to a vanishing interval
 for instanton transition (if measurements are performed with perfect resolution).
 
 Note also that
while the shrinking of topological modes (\ref{shrinking}) follows from  
Yang-Mills theory, explaining the observed  correlation of the topological modes
with the lattice strings is beyond the scope of field theory
so far.
Probably, clues are provided by theory of the defects in the dual,
string formulation but there has been no discussion of the issue in the literature.

\subsection{Protected and unprotected matrix elements}

There is  a drastic difference between results of measurements of, say, 
topological susceptibility (\ref{topologicalcharge}) and of the instanton 
size. While the value of (\ref{topologicalcharge}) does not depend on
the resolution, or lattice spacing $a$, the size 
of topological excitations changes drastically :
\beq\label{crazy}
 \big(size\big)_{resolution~a}~\sim~\exp (- const/g^{2}(a)) \cdot 
 \big(size\big)_{resolution~\Lambda_{QCD}}~,
\end{equation}
where $g^{2}(a)$ is the bare coupling.

Clearly, one cannot think of deriving (\ref{crazy}) in perturbation theory.
Rather, one should think in terms of theory of measurements \cite{arkady,adriano}.
As a matter of fact, the chiral condensate does
not depend on the resolution and can be measured either without
or with cooling while the  size of topologically non-trivial regions of gluon field
depends power-like on the resolution. 
One can talk, therefore, about `protected' and `unprotected' matrix elements.

The question is then whether we can  judge theoretically which matrix 
element is protected. Without trying to formulate here a general recipe, turn
to  examples.

The {\it local} matrix element  (\ref{bankscasher})
is given by 
\beq\label{protected}
\langle \bar{q}q \rangle ~\sim~\lim_{m_{q\to 0}}m_{q}\int
\rho(\lambda)~{d\lambda\over (\lambda^{2}+m_{q}^{2})}~~,
\end{equation}
where $\lambda$ is the eigenvalue, see (\ref{lambda}),
and $\rho(\lambda)$ is the density of states. With lattice spacing
$a\to 0$ the number of eigenfunctions  grows power-like since 
$\lambda_{max}\sim 1/a$.  However, it is clear that all the modes 
with $\lambda\gg m_{q}$ are canceled from (\ref{protected})  in the limit of the vanishing quark mass, $m_{q}\to 0$.
Thus, quark condensate gives an example of a protected matrix element.

To define instanton size in terms of a matrix element one can try a  
{\it non-local} generalization of (\ref{protected}).  Namely, consider the correlator 
\footnote{This definition, in its gauge invariant versions, see below, was suggested to me by V.A. Rubakov.} :
\begin{equation*}\langle ~\bar{q}^{\alpha}(x), q^{\beta}(0)~ \rangle~
 \end{equation*}
 where the spinor indices are contracted the same way as in
 case of local condensate (\ref{protected}) and $\alpha,\beta$ are color indices. 
Because of chirality conservation, the nonlocal correlator is not 
 contributed
 by perturbation theory and determined by instantons. 
 
 True, our  correlator  is not acceptable yet because 
 it is not gauge invariant. To amend this, one can insert  the phase factor:
 \beq\label{nonlocal}
 \langle ~\bar{q}^{\alpha}(x)\Phi_{\alpha,\beta}(x,0)q^{\beta}(0)~ \rangle~
 \equiv~f_{1}(x^{2})~~,
 \end{equation}
 where, as usual,  $\Phi(x.0)~=~P\exp i\int A_{\mu}dx_{\mu}$.
 In the quasiclassical approximation, instantons do contribute to
 (\ref{nonlocal}) and produce a non-trivial $x^{2}$ dependence
 with characteristic scale of $\Lambda_{QCD}$. This scale could be considered
 as a general definition of the size of topological excitations in terms of 
 a gauge invariant matrix element.
 Another possibility is to introduce a color scalar field $\phi_{\alpha}$
 and consider correlator of colorless spinor currents:
 \beq\label{nonlocal1}
 \langle ~\bar{q}^{\alpha}(x)\bar{\phi}_{\alpha }(x),
 q^{\beta}(0)\phi_{\beta}(0)~ \rangle~
 \equiv~f_{2 }(x^{2})~~.
 \end{equation}
An advantage of this definition is that one does not need to fix the path of integration
in the P-exponent (\ref{nonlocal}).

The question is now, whether the matrix elements (\ref{nonlocal}), 
(\ref{nonlocal1}) are protected or not.
 It is obvious that the matrix element (\ref{nonlocal}) changes greatly if measurements are 
performed with resolution of order $a$. Indeed the P-exponent entering
 (\ref{nonlocal}) has ultraviolet divergent action, the same as, say, Wilson
 line \footnote{I have learned the argument from Ph. de Forcrand, 
 in another context.} . Therefore, 
 \beq
 f_{1}(x)~\sim~\exp \big(-const{|x|\over a\alpha_{s}}\big)~~,
 \end{equation}
 and we conclude that the matrix element (\ref{nonlocal}) is not protected against
 huge resolution-dependent corrections.
 
 Turn now to another possibility, that is correlator (\ref{nonlocal1}).
 It is again very sensitive to the resolution
 (provided by the lattice spacing $a$).
  Indeed, the scalar particle acquires a quadratically divergent mass,
 \begin{equation}
 M^{2}_{\phi}~\sim~\alpha_{s}/a^{2}~,
 \end{equation}
 and the correlator   (\ref{nonlocal1}) vanishes
 at distances of order $a/\alpha_{s}$
  Of course, one could try to renormalize mass but then many orders of perturbation theory should be 
taken into account and correlator (\ref{nonlocal1})
  is no simpler than (\ref{nonlocal}) discussed above.
  
  \subsection{Unitarity}
   Thus, the correlators (\ref{nonlocal}), (\ref{nonlocal1}) are very sensitive to the resolution and subtractions.
To get rid of the problem of divergences consider first  the quenched approximation when both
  the spinor and scalar fields are non-dynamical external probes.
To learn more on the resolution dependence, turn again to the unitaruty constraints.
Inserting a complete set of intermediate states:
  \begin{equation}
   f_{2 }(x^{2})~\sim~\Sigma_{n}\exp(-m_{n}\sqrt{x^2})~~,
   \end{equation}
   where $m_{n}$ are masses of spin-1/2 `hadrons' in case of pure
   gluodynamics. Assuming  that   there 
are no such bound states of gluons, we come to
conclusion that
\begin{equation}
 f_{2 }(x^{2})~\sim~\delta({x^{2}})~~,
 \end{equation}
 which is equivalent to derivation of shrinking of the fermionic
 topological modes to a vanishing 4d volume in the continuum limit.
 
  Note, however, that in principle one cannot rule out that there exist
 spin-1/2 states in pure gluodynamics (Skyrme states). 
 Then the proof of the
 shrinking is lacking. 
 More realistically, if the fermions are treated dynamically
    there are spin-1/2 colorless states.  The width of the $f_{2 }(x^{2})$
   distribution
   is controlled then  by the corresponding masses.

   Thus, there is a hint that in the unquenched approximation
   the topological modes might not shrink any longer to a vanishing 4d
   volume. Moreover, our derivation of the shrinking given above and 
   based on consideration of the correlator of topological charge densities
   also utilizes the quenched approximation. Indeed, we assumed
   a local correspondence between integrals over topological modes and bumps of
   topological charge density. Which is true only in the quenched approximation.
  
   \section{Localization of scalars}
   \subsection{Quasiclassical picture}
   
   After experience with fermions, it is only natural to study localization
   properties of test color scalar particles in Yang-Mills vacuum \cite{scalars}.
   To this end, one considers solutions of the equation
   \begin{equation}\label{scalars}
   D^{2}\phi_{\lambda}~=~\lambda^{2}\phi_{\lambda}  ~~,
   \end{equation}
   where the covariant derivative is constructed on the vacuum-field configurations
   $\{A_{\mu}^{a}\}$. The analysis 
   has been performed for SU(2) case only but for various values of the color
   spin, T=1/2,1,3/2.
   
   For the purpose of orientation let us begin with the instanton case
   where the equation (\ref{scalars}) can be solved analytically
   \cite{thooft}. Moreover, let us assume that the momentum 
   of the scalar is much larger
   than the inverse instanton size, $p^{2}\gg \rho^{-2}$.
   Then the effect of interaction with the external instanton field is a
   mass shift of the scalar:
   \begin{equation}\label{tachyonic}
   p^{2}~~\rightarrow~~p^{2} ~-~T(T+1)\rho^{-2}~~,
   \end{equation}
   where $p^{2}$ is the momentum squared of a free particle.
   Eq. (\ref{tachyonic}) looks like introduction of a tachyonic mass.
   But this is true only as an expansion at large $p^{2}$.
   There are no actual tachyonic states since
   all the eigenvalues of the  Eq. (\ref{scalars})
   are positive definite. 
   
   \subsection{Hard fields, tachyonic mass}
   
  If   hard, or original vacuum fields, $A_{\mu}\sim1/a$,  are used to evaluate the eigenvalues
  $\lambda^{2 }$ for scalar particles,
  then there is an  ultraviolet divergent
  radiative mass correction,   \begin{equation}
   \delta M^{2}~\sim~ \alpha_{s}a^{{-2}}~.
   \end{equation}
On the lattice, indeed, the minimum
   eigenvalue $\lambda^{2}_{min}\sim a^{{-2}}$. 
   According to the standard perturbative prescription, one is free to renormalize
   the mass to any physical value by subtractions.
   For example, subtraction constant $M^{2}=-\lambda_{min}^{2}$ 
   would correspond to zero renormalized mass.
   After fixing the renormalized mass, the theory is fully determined
   perturbatively.
   
   Non-perturbative treatment of the problem on the lattice \cite{scalars}
   brings results  different from the text-book pattern.
   Namely, there turn to be two candidates for identification with the
   radiative  mass correction.
   Apart from $\lambda_{min}^{2}$ there is another remarkable 
   eigenvalue, $\lambda^{2}_{mob}$ such that for 
   \begin{equation}\label{interval}
   \lambda^{2}_{min}~<~\lambda^{2}~<
   \lambda^{2}_{mob}
   \end{equation}
   the corresponding eigenfunctions are localized on  finite volumes
\footnote{There exist various detailed  mechanisms of
 localization.  The observed pattern of localization of the scalars
 does not correspond to the Anderson localization.}..
   If we identify the mobility edge with the radiative mass
   and introduce subtraction term $M^{2}=-\lambda_{mob}^{2}$,
     then there is an advantage that higher eigenvalues $\lambda_{n}^{2}>\lambda_{mob}^{2}$ might be associated with
     standard plane waves. 
   However, the
   renormalized eigenvalue 
      $$\tilde{\lambda }^{2}~\equiv~\lambda^{2}~-~\lambda^{2}_{mob}$$
   is then negative in the interval (\ref{interval}).
   Tachyonic states are  becoming reality.
   
   More generally, existence of the two  scalars, $\lambda^{2}_{mob}$
   and $\lambda^{2}_{min}$ associated with  a single particle defies the standard 
   classification scheme of states with respect to the Poincare group.
   
   \subsection{Further critical exponents}
   
   There are new type of states, localized  states, and  
   new critical exponents can be introduced \cite{scalars}:
   \begin{equation}
   \lambda^{2}_{mob}~-~\lambda^{2}_{min}~\sim~\Lambda^{2}_{{QCD}}
   (a\cdot \Lambda_{QCD})^{-\alpha}~~~,~~~
   V_{loc}(\lambda_{{min}})~\sim~\Lambda_{QCD}^{-4}(a\cdot\Lambda_{QCD})^{\beta}
   ~,
   \end{equation}
   where $V_{loc}(\lambda_{min})$ is the localization volume.
   
   The results for  $\alpha,\beta$ turn  once again absolutely unexpected. Namely, 
the values of $\alpha,\beta$ depend crucially on the
   color spin:
   \begin{eqnarray}\label{results}
   \alpha(T=1/2)~~\approx~~0~,~~~\beta(T=1/2)~~\approx~~0~;\\
    \alpha(T=1)~~\approx~~1~,~~~\beta(T=1)~~\approx~~2~\nonumber.
    \end{eqnarray}
   In other words, the interval (\ref{interval}) appears to be divergent
   in the continuum limit for the adjoint case. Then the renormalization program
   for the scalar particles in the adjoint representation   cannot actually
   be performed \cite{scalars}. 
   
   \subsection{Preludes to interpretations}
   There has been no detailed theoretical discussion of the data \cite{scalars}
   on localization of scalar particles
   in the literature and we can add only straightforward remarks.
   
   The measurements brought results which could not
   be foreseen perturbatively.  
   What is the basic difference between the two approaches,
   perturbative and non-perturbative?
   Non-perturbatively, or on the lattice equations of motion are 
   not valid for any particular vacuum-fields configuration $\{A_{\mu}^{a}\}$:   
\begin{equation}\label{currents}
   D_{\mu}G_{\mu\nu} ~=~`j_{\nu}~'~~,~~~|`j_{\nu}~'|~\sim~a^{-3}~~, 
    \end{equation}
   with a non-vanishing `current' $j_{\nu}$. 
The Feynman graphs, on the other hand,
   are equivalent to evaluating matrix elements
   and equations of motion are (imposed to be)  true.
   
   Of course, one is tempted to disregard the spurious currents (\ref{currents}).
   However, it does not seem wise to throw away these currents
   in the Yang-Mills case. Indeed, confinement is nothing else but
   a manifestation of instability of the perturbative vacuum
   and the currents (\ref{currents}) provide original fluctuations
   which allow the system to learn about the non-perturbative
   instability. Killing all the fluctuations (\ref{currents}) on the lattice-spacing
   scale would extinguish confinement on the $\Lambda_{{QCD}}$ scale.
   
   Mostly, the fluctuations on the lattice-spacing scale (\ref{currents}) 
   are dying without producing any effect but sometimes they develop
   into long-range structures. It is like in the Universe, original fluctuations
   of the density of matter at short distances develop into huge voids
   and clusters at later stages.
   
The instability of the perturbative vacuum is quantum-mechanical in nature
and is revealed through barrier tyransitions.
This implies that
    regions of strong instabilities are
   allowed to occupy only a vanishing part of the whole volume,
   \begin{equation}
   V_{{instab}}~\sim~\exp(-const/g^{2}(a))V_{tot}~\sim~(a\cdot \Lambda_{QCD})^{\rho}V_{tot}
  ~~.
  \end{equation} 
   Lattice strings and chiral defects considered above 
   are  example of such regions of instability corresponding to
   $$\rho_{strings}~=~2~~,~~\rho_{chiral}~=~3~~,$$
   respectively.
   
   Fluctuations off the perturbative vacuum provide a random component
   of the non-Abelian fields. Stochastic fields, in turn, provide localization.
   Scalar particles are better tools to study localization since for 
   fermions the gyroscope effect resists the localization strongly.
   Thus, studies of localization and confinement are closely related to each other.
   
 Measurements \cite{scalars} reveal a  variety of stochastic fields
on different scales:
\begin{eqnarray}
spin ~T=1/2~is ~sensitive ~to~the~\Lambda_{QCD}~scale;\\
spin~ T=1~ is ~sensitive ~to~ the ~\sqrt{\Lambda_{QCD}\cdot a} ~scale;\\
spin~T=3/2 ~ is~ sensitive~to~the ~a~scale.
\end{eqnarray}    
Note that stochastic fields on the
    scale of $\Lambda_{QCD}$ are commonly discussed.
The appearence of the other scales is new.
The data reveal also a kind of threshold behaviour
of the dependence on 
 the value of the coupling of scalars to gluons. 
For example, stochastic fields on the mixed scale of $\sqrt{\Lambda_{QCD}\cdot a}$
are strong enough for binding spin T=1 scalars but the spin T=1/2 escapes from
that scale and get bound only on the scale of $\Lambda_{QCD}$.
 In a way, we are lucky to have such simple means to probe various scales,
by varying the color spin. There is no simple explanation of the phenomenon, however.
   
   There is little doubt that we are only at the very beginning
   of learning the stochastic fields in the vacuum. 

     \section{Conclusions}
  
 Lattice measurements on original, hard fields $A_{\mu}^{a}\sim 1/a$
 brought unexpected, 
 power-like dependences on the lattice spacing
 \cite{ultraviolet,3d,random,scalars}.
 Generically, one can talk about low-dimensional defects in vacuum
 \cite{vz5}.  Theory of such defects is in its infancy.
  In this talk we presented a few observations
  on power-like dependences on the lattice spacing, which are
  not necessarily closely related to each other.  
  
  In particular, we argued that the lattice spacing plays the role of resolution
  \cite{arkady}.  A simple analogy from quantum mechanics is measurements of instantaneous
  velocity of a particle: it tends to infinity with space resolution tending
  to zero.
  
  More specifically, in case of topological fermionic modes (in the
  quenched approximation) one can utilize dispersion relations 
  to argue that  the volume occupied by the modes vanishes
  indeed in the continuum limit. The physics behind is that large
  fluctuations of the topological density are due to under-the-barrier
  transitions in the Euclidean space. Therefore, they do not correspond
  to any physical intermediate states and are described by local,
  or subtraction terms in field theory \cite{arkady}.
  
  Thus, chirality-related defects seem to be the first case 
  when shrinking to a vanishing (in the continuum limit)
  submanifold of the whole 4d space
  can be understood and predicted within standard formulation of
  field theory. Moreover, lattice measurements are always
  restricted to some finite values of the lattice spacing
  and one could suspect that 'lower-dimensional defects'
  are in fact pre-asymptotic. The field -theoretic derivation
  of existence of a low-dimensional defect has an advantage of
  holding directly in the continuum limit. 
  
  Another, and probably most interesting aspect is that dual formulations
  of YM theories with large $N_{c}$ predict existence of low-dimensional
  vacuum defects. One of the earliest of such predictions is existence
  of domain walls, or 3d  defects related to topology \cite{witten3}.
  Although the argumentation does not apply literally if $N_{c}$ is not large,
  the basic geometrical constructions could survive in the $N_c=2,3$ 
   cases as well.
  Then, to uncover low-dimensional defects one needs measurements
  with high resolution since only explicit lattice dependence distinguishes, say,
  3d defects from 4d excitations.
 
 As for the scalars, let me mention in conclusion only a single point.
   At this conference a lot of attention is devoted
   to hypothetical new forms of matter, due to scalar fields, see, e.g.,
   \cite{starobinski}. The lattice simulations indicate strongly that one can,
   indeed, expect unusual states for scalar fields which look like
   localized states when continued to the Euclidean space. I am not sure that
   specifically such states have ever been discussed in the literature 
   on new forms of matter.

  \subsection*{Acknowledgments}
  I am thankful to the organizers of the International seminar
  ``Quarks06'' for the invitation and hospitality.
  
  Observations presented in this talk were partly elaborated in collaboration
  with A.I. Vainshtein \cite{arkady}. I am thankful to A. Di Giacomo, 
  Ph. de Forcrand,
  J. Greensite and V.A. Rubakov for
  useful discussions. Special thanks are to members of the ITEP Lattice Group,
  F.V. Gubarev, A.V. Kovalenko, S.M. Morozov, M.I. Polikarpov, S.N. Syritsyn
  for letting me know their results and for thorough discussions.


\begin{thebibliography}{99}
\bibitem{shuryak}
Th. Schafer, E.V. Shuryak,
{\it ``Instantons in QCD''}, 
{\it Rev. Mod. Phys.} {\bf 70} (1998) 323, [arXiv:hep-ph/9610451].


\bibitem{teper1}
M. Teper,
 {\it`` Topology in QCD''},
{\it Nucl. Phys. Proc. Suppl.} {\bf 83} (2000)  146, [arXiv:hep-lat/9909124].

\bibitem{witten3}
E. Witten, 
{\it`` Instantons, The Quark Model, And The 1/N Expansion''},
{\it Nucl. Phys.}, {\bf B145} (1978) 110;
{\it`` Theta dependence in the large N limit of four-dimensional gauge theories''},
{\it Phys. Rev. Lett.}  {\bf 81} (1998) 2862,
[arXiv:hep-th/9807109].

\bibitem{horvath}
 I. Horvath, et al.,
 {\it ``On the local structure of topological charge fluctuations in QCD.''}, {\it Phys. Rev.} {\bf D67} (2003),
 011501, [arXiv:hep-lat/0203027].
 
\bibitem{greensite}
J. Greensite, 
{\it `` The Confinement problem in lattice gauge theory''},
{\it Progr. Part. Nucl. Phys.} {\bf 51} (2003) 1,
[arXiv:hep-lat/0301023].

\bibitem{ultraviolet}
F.~V.~Gubarev et al.,
{\it `` Fine tuned vortices in lattice SU(2) gluodynamics''},  {\it Phys. Lett.} {\bf B574} (2003) 136, [arXiv:hep-lat/0212003];\\
V.G. Bornyakov. et al.,
 {\it `` Anatomy of the lattice magnetic monopoles''}, 
 {\it  Phys. Lett.} {\bf B537} (2002) 291, [arXiv:hep-lat/0103032].



\bibitem{vz5}
 V.I. Zakharov,
 {\it``Lower-dimensional vacuum defects in lattice Yang-Mills theory''}, 
 {\it Yad. Fiz.} {\bf 68} (2005) 603, [arXiv:hep-ph/0410034];
{\it``  Dual string from lattice Yang-Mills theory''},
{\it  AIP Conf. Proc.} {\bf 756} (2005) 182, [arXiv:hep-ph/0501011].



\bibitem{3d}
 A.V.~Kovalenko, M.I.~Polikarpov, S.N.~Syritsyn,
V.I.~Zakharov, 
{\it `` Three dimensional vacuum domains in four dimensional SU(2) 
gluodynamics''}, {\it Phys.Lett.} {\bf B613} (2005) 52,
[arXiv:hep-lat/0408014 ];\\
M.I. Polikarpov, S.N. Syritsyn , V.I. Zakharov,
{\it ``  A novel probe of the vacuum of the lattice gluodynamics''},
 {\it JETP Lett.} {\bf 81} (2005) 143, [arXiv:hep-lat/0402018]. 

\bibitem{random}
C. Aubin, {\it et al.} {\it ``The Scaling Dimension of
Low Lying Dirac Eigenmodes And Of The Topological Charge Density''},
{\it  Nucl. Phys. Proc. Suppl.} {\bf 140} (2005) 626,
[arXiv:hep-lat/0410024];\\
F.V. Gubarev, S.M. Morozov, M.I. Polikarpov, V.I. Zakharov,
{\it `` Localization of low lying Eigenmodes for chirally symmetric
Dirac operator''}, 
{\it JETP Lett.}{ \bf 82} {343}(2005), [arXiv: hep-lat/0505016];\\
Y. Koma {\it et al.},{\it ``Localization properties of the topological
charge density and the low lying eigenmodes of overlap fermions''},
PoS LAT2005:300,2005 [arXiv:hep-lat/0509164];\\
C. Bernard {\it et al.}, {\it ``More evidence of localization in
low-lying Dirac spectrum''} PoS LAT2005:299,2005 [arXiv:hep-lat/0510025];\\
V. Weinberg et al.,
{\it ``The QCD vacuum probed by overlap fermions''},
[arXiv:hep-lat/0610087]. 

\bibitem{neuberger}
R.~Narayanan, H.~Neuberger,
{\it`` A Simulation of the Schwinger model in the overlap formalism''},
{\it Nucl.\ Phys.\ B} {\bf 443}, (1995) 305,
[arXiv:hep-th/9411108];\\
H.~Neuberger, 
{\it `` Exactly massless quarks on the lattice''},
{\it Phys. Lett.} {\bf B417} (1998) 141, [arXiv:hep-lat/9707022];
{\it  `` More about exactly massless quarks on the lattice''},
{\it Phys. Lett.} {\bf B427} (1998) 353, [arXiv:hep-lat/9801031].

\bibitem{gubarev}
P.Yu. Boyko, F.V. Gubarev, S.M. Morozov, 
{\it`` SU(2) gluodynamics and HP1 sigma-model embedding: Scaling, topology and confinement''},
{\it Phys. Rev.}
{\bf D73} (2006) 014512, [arXiv:hep-lat/0511050];\\
P.Yu. Boyko, F.V. Gubarev, S.M. Morozov,
{\it ``SU(2) gluodynamics and HP1 sigma-model embedding: Scaling, topology and confinement''},
 {\it Phys. Rev.} {\bf D73}  (2006) 014512 [arXiv:hep-lat/0511050]. 

 \bibitem{correlator}
A.V. Kovalenko, S.M. Morozov, M.I. Polikarpov , V.I. Zakharov,
 {\it ``On topological properties of vacuum defects in lattice Yang-Mills theories''}, 
 [arXiv:hep-lat/0512036].

\bibitem{arkady}
A.I. Vainshtein, V.I. Zakharov, in preparation. 

\bibitem{adriano}
V.I. Zakharov,
 {\it``Matter of resolution: From quasiclassics to fine tuning''}, 
 in {\it``Sense of Beauty in Physics. A volume in honour of Adriano Di Giacomo''},
 Pisa University Press (2006), p 61,
 [arXiv:hep-ph/0602141].


\bibitem{seiler}
M. Aguado, E. Seiler, 
{\it `` The Clash of positivities in topological density correlators''}, {\it Phys. Rev.} {\bf D72} (2005) 094502,
[arXiv:hep-lat/0503015].

\bibitem{thacker}
H.B. Thacker,
{\it``D-branes and topological charge in QCD''}, 
{\t PoS LAT2005} (2006) 324, [arXiv:hep-lat/0509057].


 \bibitem{scalars}
J. Greensite et al., 
 {\it ``Localized eigenmodes of covariant Laplacians in the Yang-Mills vacuum''},
 {\it Phys. Rev.} {\bf D71} (2005) 114507, [arXiv:hep-lat/0504008];\\
J. Greensite et al., 
 {\it `` Peculiarities in the spectrum of the adjoint scalar kinetic operator in Yang-Mills theory.''},  {\it Phys. Rev.} 
 {\bf D74} (2006) 094507,  [arXiv:hep-lat/0606008].
 
  \bibitem{thooft}
  G. 't Hooft
{\it`` Computation Of The Quantum Effects Due To A Four-Dimensional Pseudoparticle''}, {\it  Phys. Rev.} {\bf D14} (1976) 3432.
 
 \bibitem{starobinski}
 R. Gannouji, D. Polarski, A. Ranquet , A.A. Starobinsky,
 {\it``Scalar-Tensor Models of Normal and Phantom Dark Energy''},
 {\it JCAP} {\bf 016} (2006)  0609, [arXiv:astro-ph/0606287];\\
 V.A. Rubakov, 
  {\it`` Phantom without UV pathology''}, [arXiv:hep-th/0604153]. 
\end{thebibliography}
\end{document}